%% file: paper.tex
\documentclass[sigconf,authorversion]{acmart}

\usepackage{tabularx}

\newcolumntype{R}[1]{>{\raggedleft\arraybackslash}p{#1}}

\AtBeginDocument{%
  \providecommand\BibTeX{{%
    \normalfont B\kern-0.5em{\scshape i\kern-0.25em b}\kern-0.8em\TeX}}}

\setcopyright{acmcopyright}
\copyrightyear{2021}
\acmYear{2021}
\acmDOI{10.1145/1122445.1122456}

\acmConference[SIGCSE TS '22]{SIGCSE TS '22: The 53rd ACM Technical Symposium on Computer Science Education}{March 02--05, 2022}{Providence, RI, USA}
\acmBooktitle{SIGCSE TS '22: The 53rd ACM Technical Symposium on Computer Science Education, March 02--05, 2022, Providence, RI, USA}
\acmPrice{15.00}
\acmISBN{978-1-4503-XXXX-X/18/06}

\newcommand{\DEERS}{DEERS}

\newcommand{\DEERSWebsite}{http://empiricalcsed.org}

\begin{document}

\title{Training Computing Educators to Become Computing Education Researchers}

\author{Jeffrey C. Carver}
\orcid{0000-0002-7824-9151}
\affiliation{%
 \institution{University of Alabama}
 \country{USA}
}
\email{carver@cs.ua.edu}

\author{Sarah Heckman}
\orcid{0000-0003-4351-8611}
\affiliation{%
 \institution{North Carolina State University}
 \country{USA}
}
\email{sarah_heckman@ncsu.edu}

\author{Mark Sherriff}
\orcid{0000-0002-1745-205X}
\affiliation{%
 \institution{University of Virginia}
 \country{USA}
}
\email{sherriff@virginia.edu}

\authornote{All authors contributed equally to this work.}

\begin{abstract}

\input{section-abstract}

\end{abstract}



\maketitle

\section{Introduction}
\label{introduction}
\input{section-introduction}

\section{Background}
\label{background}
\input{section-background}

\section{\DEERS}
\label{workshop}
\input{section-workshop}

\section{Evolution to Online}
\label{evolution}
\input{section-evolution}

\section{Reflection \& Lessons Learned}
\label{reflection}
\input{section-reflection}

\section{Alumni Response}
\label{alumni}
\input{section-alumni}

\section{Future Directions}
\label{future}
\input{section-future}

\begin{acks}
\input{section-acknowledgements}
\end{acks}

\bibliographystyle{ACM-Reference-Format}
\bibliography{references}

\end{document}

%% file: section-abstract.tex
The computing education community endeavors to consistently move forward, improving the educational experience of our students.
As new innovations in computing education practice are learned and shared, however, these papers may not exhibit the desired qualities that move simple experience reports to true Scholarship of Teaching and Learning (SoTL).
We report on our six years of experience in running professional development for computing educators in empirical research methods for social and behavioral studies in the classroom.
Our goal is to have a direct impact on instructors who are in the beginning stages of transitioning their educational innovations from anecdotal to empirical results that can be replicated by instructors at other institutions.
To achieve this, we created a year-long mentoring experience, beginning with a multi-day workshop on empirical research methods during the summer, followed by regular mentoring sessions with participants, and culminating in a follow-up session at the following year's SIGCSE Technical Symposium.
From survey results and as evidenced by eventual research results and publications from participants, we believe that our method of structuring empirical research professional development was successful and could be a model for similar programs in other areas.

%% file: section-introduction.tex
Members of the computing education community invest a large amount of effort developing new methods to improve the educational experience of their students.
These educators hope by reporting these experiences in conference and journal venues, other members of the computing education community will adopt, or attempt to adopt, their methods into their own environments.
This approach has worked quite well for some educational innovations.
Because of the potential for widespread adoption and the impact that such adoption could have on future generations of computer science students, it is important for these research reports to include adequate evidence that the educational innovation is effective at achieving its stated goal.

Excellent educators already reflect on their teaching experience and actively seek to improve from semester to semester.
Sometimes they share these reflections with the larger computing education community in the form of the experiences described above.
By increasing the rigor of the analysis of their classroom interventions an educator can move from reflective teaching to the \textit{Scholarship of Teaching and Learning} (SoTL)~\cite{Bishop-Dietz:12}.
SoTL includes: asking appropriate research questions, designing and conducting studies to answer those questions, applying appropriate analyses to the results, providing details about where the work fits into the existing body of knowledge, and describing how to interpret the results in light of the threats to validity~\cite{Bishop-Dietz:12,Fincher-Petre:04}.

Many of the reports on excellent teaching practices in the computing education community may meet the standards of scholarly work~\cite{Bishop-Dietz:12}, which includes items like novelty, enough detail for replication, and being peer reviewed.
However, much of the work lacks rigor, including missing key items like appropriate study methods, related work, and threats to validity~\cite{Bishop-Dietz:12, valentine2004cs, randolph2008methodological,al2016updated}.
Recent reviews of the computing education literature show that while there is empiricism in computing education, the literature is lacking the rigor appropriate for SoTL~\cite{Bishop-Dietz:12,valentine2004cs,randolph2008methodological,al2016updated,heckman2021systematic}.
These shortcomings slow the progress of SoTL within the computing education research (CER) community.

To help remedy this situation, we developed a program called \textit{Designing Empirical Education Research Studies} (\DEERS) to train Computer Science educators on the concepts of human-based empirical research.
\DEERS~includes a intense summer cohort workshop followed by a year of one-on-one mentoring.
Our experience has shown that many in the computing education community conducted their dissertation research on topics that did not involve human subjects (e.g. algorithms or networks).
Therefore, educators often need to learn the concepts involved in conducting computing education research, which is heavily human-based.
\pagebreak

The goals of \DEERS~are:

\begin{itemize}
    \item To introduce computing educators to the concepts of human-based empirical research
    \item To help computing educators define their research questions and study design in a way that will be most helpful to the larger community
    \item To mentor computing educators through execution of their study to publication of the research report
    \item To increase the level of sophistication about human-based empirical research throughout the computing education community
\end{itemize}

%% file: section-background.tex
The design of \DEERS~was heavily influenced by literature related to SoTL and to computing education research.  
We utilized many of the resources below, along with our backgrounds, 
in the development and evolution of \DEERS.

\subsection{Scholarship of Teaching and Learning}
SoTL is the application of empiricism to computing education research.
We define empiricism as ``validation based on observation of an intervention''.
An empirical validation draws conclusions based upon observed evidence rather than argumentation, proof, or some other means~\cite{shaw2003writing}.
Empirical validation does not solely involve experimentation or the scientific method, but incorporates the ``method of science''~\cite{Fincher-Petre:04}.
The method of science considers both inductive and deductive paradigms for gathering evidence to answer a research question~\cite{Fincher-Petre:04}.
An increase in empiricism may help move CER to a recognized sub-area of computer science~\cite{Clear:2006:VCS:1315803.1315806}.

\subsection{Computing Education Research}
A large portion of the computing education literature includes scholarly work on excellent teaching practice~\cite{Bishop-Dietz:12}. 
These experience reports are novel, include details for adoption, and are peer reviewed (like this paper).
While experience reports are valuable and describe an ``educational approach or tool, the context of use, and provide a rich reflection on what did or didn't work, and why,''\footnote{From the SIGCSE Technical Symposium 2022 call for papers.} more rigorous research that directly assesses a research question, can move the field of computing education from reflective teaching to SoTL providing the foundation for computing-specific educational theory~\cite{Fincher-Petre:04}.

However, much of published CER work lacks reporting of elements appropriate for research, suggesting the need for additional work to support computing education practitioners and researchers in development of research studies and reporting the results of those studies. 
Reviews of computing education literature have found gaps in reporting of research questions~\cite{randolph2008methodological, heckman2021systematic, lishinski2016methodological}, related work, lack of detail in the methods~\cite{ihantola2015educational, luxton2018introductory}, study context and participants~\cite{randolph2008methodological, mcgill2018improving}, threats to validity~\cite{ihantola2015educational, al2016updated}, research ethics~\cite{ihantola2015educational}, and connection of results to theories (when appropriate)~\cite{sheard2009analysis, malmi2010characterizing}.  
However, these studies do not necessarily make a distinction between research and experience reports in their findings~\cite{heckman2018SIGCSEpaper}.

An additional challenge is that replication is lacking in CER, partially due to challenges in reporting and to a bias towards original work~\cite{Ahadi2016Koli,mcgill-koli2019-replication}. 
A literature review on educational data mining by \citet{ihantola2015educational} found that only 5 studies (7\% of their reviewed papers) were replications.
A broader literature review, specifically on replication in CER by \citet{Hao:2019TOCE:Replication}, found only 2.38\% of 2,269 articles were considered replications, with only 63\% of those being successful.  
Further, many research studies do not contain a point of comparison~\cite{al2016updated,heckman2021systematic}.  Additionally, studies will frequently create new measurement instruments rather than utilize common and validated instruments~\cite{margulieux2019measurements}.

Educational research organizations provide guidelines to support the development of empirical studies and reporting on educational research~\cite{WWC2020, Schulzc-BMJ2010-ConsortStatement, apa-jars, area-empirical-social-science-2006}.  
Additional recommendations on high quality reporting, and therefore high quality research design, come from the CER community~\cite{daniels2012models, mcgill-fie2018-cer-repository,mcgill2018improving, ihantola2015educational, heckman2021systematic} and related computing fields like software engineering~\cite{runeson2009guidelines, jedlitschka2005reporting, kitchenham2008evaluating, Carver-RESER2010-ReportingGuidelines}.

%% file: section-workshop.tex
We have conducted \DEERS~ annually from 2016 through 2021. 
We plan to continue offering \DEERS~in its current form at least through the summer of 2022 and potentially longer, contingent on funding.
In our time running \DEERS, the format of the workshop has evolved as we have learned more about how to best structure a professional development experience for faculty interested in pursuing their own empirical CER projects.

\subsection{Structure and Organization}
We have offered \DEERS~six times, with four instances run in-person and two instances run fully online due to COVID.
In addition, we organized two mini-versions as workshops at the SIGCSE Technical Symposium.
\DEERS~is structured as a year-long mentoring experience, beginning with a multi-day summer workshop on empirical research methods, followed by regular one-on-one mentoring sessions with participants, and culminating in a follow-up session at the following year's SIGCSE Technical Symposium.  
Our discussion here focuses on the in-person workshop. 
We discuss the online version in Section \ref{evolution}.
Each cohort of participants begin the program with a two-and-a-half day, in-person summer workshop. 
We modeled the structure and schedule of the summer workshop after a typical short conference, including group meals and social events, such as tours of the local area.
With funding support from the NSF, participants receive a stipend to cover travel to the host institution.
We arrange for housing at a local hotel within walking distance of the workshop venue.

The schedule for the in-person workshop alternates between presentations from the organizers on aspects of empirical research, independent work where participants develop their own projects, and small group discussion where participants share and receive feedback.
Based on our observations and participant feedback, the small group discussions are one of the most valuable aspects of \DEERS.
Each group has three or four participants, one organizer, and one or more alumni of \DEERS.
Giving participants the opportunity to evaluate and provide feedback on other participants' projects can be just as valuable as receiving feedback on their own work, as they are learning to better identify the needs of empirical research projects.
We repeat this cycle of short lesson, independent work, and group discussion and feedback for each of the key steps in designing an empirical research project.
The goal is for participants to leave the workshop with a refined research question, a research plan, and feedback on any potential IRB protocol needs.

After returning to their own institutions, participants begin work on their research projects.
Throughout the academic year, mentors meet regularly with participants, offering guidance and feedback as they execute the research plan.
The cohort comes back together at the next SIGCSE Technical Symposium for a follow-on short workshop on data analysis.
At that point, participants begin plans for a paper submission for either the coming SIGCSE Technical Symposium or another conference.

\subsection{Recruiting Participants}
One of our objectives is to reach faculty that might not normally have the time or resources to pursue educational research or those that do not have much experience with social and behavioral empirical research practices.
To accomplish this goal, we begin by broadly inviting the SIGCSE community to apply to \DEERS~via the general members mailing list.
We also recruit participants through word of mouth from previous participants (after cohort 1) and by presenting our work at the NSF Showcase at the SIGCSE Technical Symposium.
The applicants propose a potential research question or topic they are interested in exploring along with their goals and motivations for applying for \DEERS.
We use this information to select a broad range of faculty and projects for each cohort and to ensure that we have the proper expertise to aid them with their projects.
Our overall acceptance rate to \DEERS~has been 45\%, including ineligible applicants - those from outside the United States (due to NSF funding restrictions) and graduate students.
Table~\ref{tab_participant_demographics} provides the demographics of the accepted participants.
Overall, 60\% were faculty at smaller, primarily teaching institutions, 17\% were teaching-track faculty at R1 universities, and 22\% were tenure-track faculty at R1 universities.
As we discuss in our future work, we hope to provide a separate experience for graduate students interested in CER.

\begin{table}[!htb]
\centering
\caption{Participant Demographics}
\label{tab_participant_demographics}
\input{table-participants}
\end{table}

\vspace{-8pt}

\subsection{Content}
We structured \DEERS~to systematically walk the participants through the creation, refinement, execution, and analysis of an empirical research project.
In the early iterations of \DEERS, we attempted to go through every aspect of a research project during the in-person workshop.
After participant feedback and realizing we needed to cover the planning aspects of empirical research deeper, we moved our content regarding data analysis to the follow-on workshop, which is when most participants were finishing collecting data.
The content and schedule below is what we use for the in-person workshops.

\subsubsection{Module 1: Scholarship of Teaching and Learning}
\begin{itemize}
    \item Length: 1 hour
    \item Learning Outcome: Understand the role of SoTL and how it differs from simply teaching
    \item Activities: Short lecture
\end{itemize}
The SoTL module helps participants understand how to move from \textit{scholarly teaching} to the \textit{scholarship of teaching and learning}. 
The module also provides an overview of \DEERS~including the other modules and mentoring over the next year.

\subsubsection{Module 2: Refining a Research Question}
\begin{itemize}
    \item Length: 3 hours
    \item Learning Outcomes: Learn the desirable aspects of a good research question and refine their own research question 
    \item Activities: Short lecture, guided worksheet, group discussion and feedback
\end{itemize}
A successful empirical research project must start with a specific and well-scoped research question.
During this session, we discuss the importance of establishing the research question before beginning the study and then give a short lesson on key aspects of a good research question: 1) interesting to the community; 2) answerable; 3) repeatable; 4) measurable; and 5) appropriately scoped.
We ask all participants to come to the first day of \DEERS~with a prospective research question.
After the presentation, the participants have independent time to refine their question before moving into small groups to share their questions and receive feedback.
While some participants merely tweak their research questions, sometimes they identify entirely new research directions during these small group conversations, as alums and fellow participants help them identify a more interesting or effective question in the same topic area.
The participants end the day with their refined research question and homework to do some background reading.

\subsubsection{Module 3: Designing a Research Study}
\begin{itemize}
    \item Length: 4 hours
    \item Learning Outcomes: Learn the details of empirical research, including variables, data types, quantitative vs. qualitative research, basic study designs, grouping participants into treatments, threats to validity, and pilot studies 
    \item Activities: Short lecture, guided worksheets, group discussion and feedback
\end{itemize}
There are many interrelated aspects important to developing a well-designed empirical study.
During this session, we begin with an overview of empirical research focused on human subjects.
We then discuss the different types of \textit{independent} and \textit{dependent} variables and how researchers should consider them in their study design.
After a discussion of \textit{qualitative} and \textit{quantitative} data types, we then describe a number of basic \textit{study designs}, highlighting the strengths and weaknesses of each one.
We conclude the session with a discussion about common threats to validity and how researches can reduce those threats during study design, where possible.
Then, starting from the research question defined in the first session, the participants have a homework assignment to develop an initial study design that identifies the key variables, lists out potential data sources, notes potential designs, and identifies threats to validity.

\subsubsection{Module 4: Ethics and the Institutional Review Board}
\begin{itemize}
    \item Length: 1 hour
    \item Learning Outcomes: Understand the ethics associated with design human-subject studies and understand the role of the Institutional Review Board (IRB) and how to interact with them at a given institution
    \item Activities: Short lecture, individual investigation of local IRB policies and procedures
\end{itemize}

For many participants in \DEERS, educational research is their first experience working with human subjects.
During this session, we discuss some of the key concepts related to designing and conducting ethical human subjects research, including those issues that arise specifically when the human subjects are students in their own classrooms.
We cover topics including \textit{risk}, \textit{beneficence}, \textit{compensation}, \textit{deception}, and \textit{debriefing}.
We then spend time discussing \textit{informed consent}, including what it is and how to ethically obtain it from participants.
Finally, we cover some of our lessons learned in working the IRBs at various institutions, including the types of documentation required and some of common issues that arise with IRB protocols.
Based on this information, the participants revisit the study design created previously to make any adjustments necessary based on considerations of ethics.

\subsubsection{Module 5: Data Collection}
\begin{itemize}
    \item Length: 3 hours
    \item Learning Outcomes: Identify data collection instruments and techniques that apply to a participant's given research question and learn how to properly gather, clean, and secure potentially identifiable information 
    \item Activities: Short lecture, guided worksheet
\end{itemize}
After designing the study, the next step is to identify and operationalize variables to answer the research question. 
This module covers considerations about whether data items are \textit{needed}, \textit{reasonable} to gather, and \textit{ethical} to gather. 
We discuss variables, including their \textit{reliability} and \textit{validity}.  
We also suggest data collection instruments, like validated attitudinal surveys, and validation techniques for other data collection instruments.  
We then discuss key considerations of data management, particularly related to privacy and confidentiality, and automation of data collection.  
Finally, there is a module on qualitative data collection that includes materials on participant observations, think-aloud studies, interviews, and surveys.  
After the presentation, the participants use a guided worksheet to create a data collection plan by 1) operationalizing the variables; 2) determining the instrumentation necessary to gather the data; and 3) developing a plan to collect the data in their study context.

\subsubsection{Module 6: Analyzing Data}
\begin{itemize}
    \item Length: 3 hours
    \item Learning Outcomes: Explore various statistical techniques that are often used in social and behavioral research 
    \item Activities: Follow-up workshop at the SIGCSE Technical Symposium
\end{itemize}
Analyzing data from human-subjects research can present challenges for researchers who are not familiar with this process.
This session, which we conduct as a follow-up private session at the SIGCSE Technical Symposium focuses on helping the participants with questions related to the analysis of their data, whether it is statistical or qualitative.
We chose to move this session out of the summer workshop into a follow-up SIGCSE Technical Symposium meeting because we learned that during the summer workshop, the participants were receiving enough information regarding research questions and study design, that it was not reasonable to add one more topic. 
In addition, many of the questions related to data analysis do not appear until researchers have actual data they are trying to analyze.
Therefore, in this session, we provide a brief overview of some of the most common statistical techniques used in human-subjects research.
We then facilitate a discussion and Q\&A among the participants to help them get their questions answered so they can proceed with the data analysis process.

%% file: table-participants.tex
\begin{tabular}{|p{.12\columnwidth}||R{.12\columnwidth}|R{.10\columnwidth}|R{.25\columnwidth}|R{.2\columnwidth}|}
\hline 
\textbf{Cohort} & \centering\arraybackslash \textbf{Female} & \centering\arraybackslash \textbf{Male} & \centering\arraybackslash \textbf{Under-represented Minority} &  \centering\arraybackslash \textbf{Minority Serving Institution} \\ \hline \hline
2016  & 3 & 9 & 2 & 2 \\ \hline
2017  & 3 & 5 & 1 & 1 \\ \hline
2018  & 3 & 11 & 1 & 1 \\ \hline
2019  & 5 & 5 & 1 & 0 \\ \hline
2020  & 6 & 4 & 0 & 0 \\ \hline
2021  & 6 & 3 & 0 & 1 \\ \hline \hline
Total & 26 & 37 & 5 & 5 \\ \hline

\end{tabular}

%% file: section-evolution.tex
For the 2020 and 2021 iterations of \DEERS, we converted the summer in-person workshop and the follow-up session at the SIGCSE Technical Symposium to an online format due to travel restrictions from COVID.
During the academic year, we already conducted our one-on-one mentoring virtually, so its format did not change.
However, many of the projects that were underway for the 2020-2021 academic year had to be adjusted due to challenges at each participant's institution.

After converting our own courses to an online format, we had a reasonable understanding of what would and would not work when we moved \DEERS~to a virtual format.
First, we knew we did not want full-day virtual sessions that tried to mirror the in-person experience.
However, we wanted to make sure the virtual sessions captured the most important aspect of the in-person version of \DEERS~- the small group discussions.
Our experience showed small group break out rooms in a virtual meeting tool can be effective if all participants are engaged.
Because \DEERS~consisted of a small group of faculty who had applied to join, we assumed we would have an engaged group.

When we were co-located, a typical session consisted of a short presentation, individual work, and group discussion for feedback.
For the online version, we recorded the presentations into 20-to-30 minute videos and hosted them on our website\footnote{\DEERSWebsite}, along with our lesson notes.
Further, we provided all of the individual worksheets to participants at the beginning of the summer workshop.
We then changed our schedule to hold only one question and answer session and one small group discussion session each day and extended the workshop to a fourth day.
The optional one-hour question and answer sessions in the morning allowed participants to join if they desired to ask questions.
The afternoon small group discussion sessions lasted two hours.
We asked the participants to review the lesson videos and worksheets before coming to the group discussion session so they could fully participate.
We began each small group discussion session with a quick review of the material in the video before going into small group breakout rooms to discuss the progress each person had made on their research project design.

Overall, we found the online model to be successful.
Creating the online course material, including videos and lecture notes, was extremely valuable, culminating in a public, self-paced course and reference for anyone interested in human-based empirical research.
The small group interactions were still generally successful, but carried with them the same drawbacks of any online group video chat interaction, including ''Zoom fatigue,'' distractions, and technical issues.
Conducting the workshop virtually did allow for participants who normally could not travel during the summer due to child care needs.
However, we did see an effect from the loss of group-building that occurs when all participants are co-located.
The online cohorts do not seem to be as cohesive a group as the in-person cohorts.
Prior workshops lead to several collaborations, 
but we have not seen the number of post-workshop collaborations with online participants.

%% file: section-reflection.tex
\subsection{Accountability}
One of the most important practices that we implemented as part of \DEERS~is our ongoing mentorship of the participants, which helps provide accountability.
During each summer workshop, we match each participant up with one of the three mentors primarily based upon mutual interest.
These mentoring relationships then continue formally for the next year (some have continued even beyond that).
Each mentor/mentee pair develops a plan for regular meetings over the course of the year. 
These meetings serve an important accountability role for the participants.
While the participants may have the best intentions of conducting their studies in the next academic year, we have seen instances where the busyness of course preparation and other tasks can push the educational research study down on the priority list.
The regular meetings help the participants prioritize their study and make slow, steady progress.
In addition to providing accountability, the meetings also provide the participants with a regular time to ask questions and receive feedback on their research questions, study design, and study progress.

\subsection{Planning}
A key aspect separating \textit{scholarly teaching} from the \textit{scholarship of teaching and learning} is planning.  
\DEERS's~focus on planning a research study is critical to developing successful research studies.  
By taking the time to determine a solid research plan with consideration of institutional context, research questions, and researcher time, we can support successful follow through.

\subsection{Time}
One of the most significant challenges in CER work is the time involved to complete a study.  
Because most studies incorporate some type of classroom intervention, the participant needs at least a semester or quarter worth of time to run the study.  
Several studies that require baselines or multiple comparisons may take one or two academic years to completely gather the data.  
Analysis and manuscript preparation takes additional time. 
One of the benefits of \DEERS~is that we can set expectations about study timelines up front and the accountability from the one-on-one mentoring supports the long timescales.  
Therefore publications from and impact of \DEERS~lag each cohort's participation by at least a year, more likely two.  
Those seeking support to offer similar experiences should set appropriate expectations about the timelines involved and the length of time before impactful output may be observed by funding agencies. 

\subsection{Redefining Success}
CER is challenging due to study time-scales, institutional contexts, lack of control, and other competing foci, especially for educators with significant teaching responsibilities.  
Not every \DEERS~participant completed their study.  
Some were unable to run their study because the class was canceled or their responsibilities changed.  
However, 37\% of alumni from the first five cohorts served as reviewers for SIGCSE Technical Symposium 2021. 
Others have reviewed for other SIGCSE Technical Symposium and CER venues.  
Their training in CER has contributed to the community through service.

\subsection{Alumni Recognition}
Many \DEERS~participants have gone on to publish on their research studies developed for the workshop~\cite{battestilli2018twostage,chattopadhyay2018cocurricularpd, edwards2018syntax,prather2019metacognition, ham2019guidedinquiry, mirza2019talitreview,basu2021strategies, neumann2021emotions, stephensmartinez2021studytactics}.  
In several cases, their participation in \DEERS~lead to additional CER work beyond their original study including successful grant proposals.

%% file: section-alumni.tex
We surveyed the participants three times about their experience - (1) before the first session of the summer workshop, (2) on the last day of the summer workshop, and (3) via a reflection survey sent to all previous cohorts in the summer of 2020.

\subsection{Pre-Workshop Survey Results}
The pre-workshop survey focused on establishing the needs of the current cohort of participants, asking questions regarding their proposed research question(s), the classes where the study would occur, and previous experience with CER.
Key results from this survey include:
\begin{itemize}
    \item Fifty percent of participants indicated they had some degree of previous experience with conducting CER work.
    \item Similarly, roughly half of the participants had previously published in a CER venue.
    \item One quarter of participants reported receiving funding to do CER work in the past.
    \item When asked what part of the research process the participants were most interested in learning more about, the answers included all of the topics we offered, with many indicating something akin to ``everything.''
    \item The most common response, however, was that participants wanted to learn more about data collection and data analysis techniques.
\end{itemize}

\subsection{Post-Workshop Survey Results}
For the post-workshop survey, we asked questions related to how the participant's proposed project evolved during the workshop, what resources they needed from us throughout the year to be successful, and what did and did not work well during the workshop.  
Nearly all participants reported that the workshop helped them better clarify and scope their project into something they felt was accomplishable.  
Similarly, nearly all participants indicated the things they most needed from us during the academic year was accountability for working on their project and assistance with understanding the proper statistical techniques needed for evaluating their results.

\subsection{Reflection Survey Results}
Our reflection survey that went to all participants from 2016-2020 contained three questions:
\begin{itemize}
    \item If the \DEERS~project (including the summer workshop, the follow-on SIGCSE meeting, and the individual mentoring) has helped your computing education research, please provide a brief description of how it has been beneficial to you. Please be as specific as possible.
    \item Please list any publications (either published, submitted, or in progress) that have occurred as a result of your participation in \DEERS. Next to each one, please indicate its status (in progress, submitted, published).
    \item Do you have any suggestions on how we could improve \DEERS~for future years?
\end{itemize}

Roughly 27\% of participants responded (15/54), spread across all previous years.
The alums reported \DEERS~had a significant impact on their projects, their motivation, and overall excitement with CER work:
\begin{itemize}
    \item \textit{"I would say that overall the three major things the program provided me were 1) accountability, 2) feedback, and 3) confidence. The accountability came from creating a timeline, wanting to contribute to an ROI to the \DEERS~program, and from the post-summer check-ins with my mentor. The feedback happened at every stage of the research process and was most helpful during conception/design of the study and during framing of the paper once data had been collected. I wouldn't have had nearly as much iteration in research question and study design without \DEERS. The confidence came from \DEERS~being an entry point into actually doing CSEd research - without the experience I would have felt like more of an outsider."}
    \item \textit{"I am extremely grateful for the summer workshop and for all the things that I learned there. Working with my advisor has been a blessing for a new faculty member like myself. I consider that when we accept a job in academia, we are blind to how to fulfill our newly adopted responsibilities. The \DEERS~workshop was instrumental in providing me with guidance on how to start and how to go from there."}
    \item \textit{"The \DEERS~project has helped advance my CSEd research in several ways. First, it gave me the opportunity to get feedback on a proposed research effort aimed at using software development assignments to improve computing students' ability to construct correct mathematical proofs. This may result in a dissertation topic for a PhD student arriving this fall.  Second, the \DEERS~workshops have given me a much deeper and more organized understanding of the various experimental methods used in social science and education research. This has made me a better researcher, but has also enabled me to provide better service in the peer review process. Finally, the workshops have given me a deeper understanding of the roles of research questions and hypotheses in guiding research, and in particular of their relationships to each other. This has made me a better researcher both within CSEd and in my other areas of interest. Perhaps most importantly, it has made me a more effective research advisor by giving me a framework within which to help my students see their research more clearly within the overall body of knowledge."}
\end{itemize}

There were a few suggestions for improvements, which included thoughts on the limitations of the online format for 2020, creating a research reading group and ideas that we are exploring in our future work.

%% file: section-future.tex
After having run \DEERS~for six years, we see multiple opportunities to move the project forward.

\subsection{Faculty/Student Pairs}
As we went through applications to join \DEERS~each year, we observed a number of graduate students who were interested in making CER a core part of their dissertation work.
Our stated purpose when we created \DEERS~was to help \textit{faculty} who normally would not have the time, resources, or mentoring necessary to start a CER project be able to do so.
As such, we tailored our material, delivery, and in-person experience for faculty and did not think it was appropriate to bring in graduate students because they would likely need more mentoring and/or there could be a conflict with their graduate advisor.

To address this observation, we plan to design and offer a modified version of \DEERS~for graduate students and their advisors.
By including both the student and the advisor, we hope to build on the existing mentoring relationship.
We will primarily target students and advisors who do not have extensive experience with human-based empirical research.

\subsection{Replication Workshops}
Another of our goals with \DEERS~was to help researchers execute human-based empirical research projects and reports that follow standard reporting norms to allow for replication of studies at other institutions.
To continue supporting replication in CER, we are exploring creating workshops specifically focused on replicating existing published CER projects.
We will choose one or two recent interesting projects from the literature and invite the investigators to come to the workshop to present their work.
Workshop participants will then identify how they can adapt the project to their own environment and begin planning their replication study with a focus on maximizing the replication space to maximize scientific gain from aggregating results~\cite{schmidt2009replication}.
One goal is to remove some of the perceived stigma of replication studies, as it is important for research to be generalized to multiple environments to establish its actual efficacy~\cite{Hao:2019TOCE:Replication, schmidt2009replication, Bishop-Dietz:12, Fincher-Petre:04}.

%% file: section-acknowledgements.tex
This material is based upon work supported by the National Science Foundation under Grant Nos. \#1525373, \#1525173, \#1525028. 
This work was approved by the IRB at all author institutions.
We would like to thank our workshop participants for their engagement and work in computing education research.

%% file: paper.bbl

\begin{thebibliography}{39}


\ifx \showCODEN    \undefined \def \showCODEN     #1{\unskip}     \fi
\ifx \showDOI      \undefined \def \showDOI       #1{#1}\fi
\ifx \showISBNx    \undefined \def \showISBNx     #1{\unskip}     \fi
\ifx \showISBNxiii \undefined \def \showISBNxiii  #1{\unskip}     \fi
\ifx \showISSN     \undefined \def \showISSN      #1{\unskip}     \fi
\ifx \showLCCN     \undefined \def \showLCCN      #1{\unskip}     \fi
\ifx \shownote     \undefined \def \shownote      #1{#1}          \fi
\ifx \showarticletitle \undefined \def \showarticletitle #1{#1}   \fi
\ifx \showURL      \undefined \def \showURL       {\relax}        \fi
\providecommand\bibfield[2]{#2}
\providecommand\bibinfo[2]{#2}
\providecommand\natexlab[1]{#1}
\providecommand\showeprint[2][]{arXiv:#2}

\bibitem[\protect\citeauthoryear{Ahadi, Hellas, Ihantola, Korhonen, and
  Petersen}{Ahadi et~al\mbox{.}}{2016}]%
        {Ahadi2016Koli}
\bibfield{author}{\bibinfo{person}{Alireza Ahadi}, \bibinfo{person}{Arto
  Hellas}, \bibinfo{person}{Petri Ihantola}, \bibinfo{person}{Ari Korhonen},
  {and} \bibinfo{person}{Andrew Petersen}.} \bibinfo{year}{2016}\natexlab{}.
\newblock \showarticletitle{Replication in Computing Education Research:
  Researcher Attitudes and Experiences}. In
  \bibinfo{booktitle}{\emph{Proceedings of the 16th Koli Calling International
  Conference on Computing Education Research}} (Koli, Finland)
  \emph{(\bibinfo{series}{Koli Calling '16})}. \bibinfo{publisher}{ACM},
  \bibinfo{address}{New York, NY, USA}, \bibinfo{pages}{2--11}.
\newblock
\showISBNx{978-1-4503-4770-9}
\urldef\tempurl%
\url{http://doi.acm.org/10.1145/2999541.2999554}
\showURL{%
\tempurl}


\bibitem[\protect\citeauthoryear{Al-Zubidy, Carver, Heckman, and
  Sherriff}{Al-Zubidy et~al\mbox{.}}{2016}]%
        {al2016updated}
\bibfield{author}{\bibinfo{person}{Ahmed Al-Zubidy},
  \bibinfo{person}{Jeffrey~C. Carver}, \bibinfo{person}{Sarah Heckman}, {and}
  \bibinfo{person}{Mark Sherriff}.} \bibinfo{year}{2016}\natexlab{}.
\newblock \showarticletitle{A (Updated) Review of Empiricism at the SIGCSE
  Technical Symposium}. In \bibinfo{booktitle}{\emph{Proceedings of the 47th
  ACM Technical Symposium on Computing Science Education}} (Memphis, Tennessee,
  USA) \emph{(\bibinfo{series}{SIGCSE '16})}. \bibinfo{publisher}{Association
  for Computing Machinery}, \bibinfo{address}{New York, NY, USA},
  \bibinfo{pages}{120--125}.
\newblock
\showISBNx{9781450336857}
\urldef\tempurl%
\url{https://doi-org.prox.lib.ncsu.edu/10.1145/2839509.2844601}
\showURL{%
\tempurl}


\bibitem[\protect\citeauthoryear{Association}{Association}{2006}]%
        {area-empirical-social-science-2006}
\bibfield{author}{\bibinfo{person}{American Educational~Research Association}.}
  \bibinfo{year}{2006}\natexlab{}.
\newblock \showarticletitle{Standard for Reporting on Empirical Social Science
  Research in AERA Publications}.
\newblock \bibinfo{journal}{\emph{Educational Researcher}}
  \bibinfo{volume}{35}, \bibinfo{number}{6} (\bibinfo{year}{2006}),
  \bibinfo{pages}{33--40}.
\newblock


\bibitem[\protect\citeauthoryear{Association}{Association}{2018}]%
        {apa-jars}
\bibfield{author}{\bibinfo{person}{American~Psychological Association}.}
  \bibinfo{year}{2018}\natexlab{}.
\newblock \bibinfo{booktitle}{\emph{Journal Reporting Standards (JARS)}}.
\newblock
\newblock
\shownote{Accessed: 2021-02-12.}


\bibitem[\protect\citeauthoryear{Basu, Heckman, and Maher}{Basu
  et~al\mbox{.}}{2021}]%
        {basu2021strategies}
\bibfield{author}{\bibinfo{person}{Debarati Basu}, \bibinfo{person}{Sarah
  Heckman}, {and} \bibinfo{person}{Mary~Lou Maher}.}
  \bibinfo{year}{2021}\natexlab{}.
\newblock \showarticletitle{Online Vs Face-to-Face Web-Development Course:
  Course Strategies, Learning, and Engagement}. In
  \bibinfo{booktitle}{\emph{Proceedings of the 52nd ACM Technical Symposium on
  Computer Science Education}}. \bibinfo{publisher}{Association for Computing
  Machinery}, \bibinfo{address}{New York, NY, USA},
  \bibinfo{pages}{1191–1197}.
\newblock
\showISBNx{9781450380621}
\urldef\tempurl%
\url{https://doi-org.prox.lib.ncsu.edu/10.1145/3408877.3432438}
\showURL{%
\tempurl}


\bibitem[\protect\citeauthoryear{Battestilli, Awasthi, and Cao}{Battestilli
  et~al\mbox{.}}{2018}]%
        {battestilli2018twostage}
\bibfield{author}{\bibinfo{person}{Lina Battestilli}, \bibinfo{person}{Apeksha
  Awasthi}, {and} \bibinfo{person}{Yingjun Cao}.}
  \bibinfo{year}{2018}\natexlab{}.
\newblock \showarticletitle{Two-Stage Programming Projects: Individual Work
  Followed by Peer Collaboration}. In \bibinfo{booktitle}{\emph{Proceedings of
  the 49th ACM Technical Symposium on Computer Science Education}} (Baltimore,
  Maryland, USA) \emph{(\bibinfo{series}{SIGCSE '18})}.
  \bibinfo{publisher}{Association for Computing Machinery},
  \bibinfo{address}{New York, NY, USA}, \bibinfo{pages}{479–484}.
\newblock
\showISBNx{9781450351034}
\urldef\tempurl%
\url{https://doi-org.prox.lib.ncsu.edu/10.1145/3159450.3159486}
\showURL{%
\tempurl}


\bibitem[\protect\citeauthoryear{Bishop-Clark and Dietz-Uhler}{Bishop-Clark and
  Dietz-Uhler}{2012}]%
        {Bishop-Dietz:12}
\bibfield{author}{\bibinfo{person}{Cathy Bishop-Clark} {and}
  \bibinfo{person}{Beth Dietz-Uhler}.} \bibinfo{year}{2012}\natexlab{}.
\newblock \bibinfo{booktitle}{\emph{Engaging in the Scholarhip of Teaching and
  Learning}}.
\newblock \bibinfo{publisher}{Stylus Publishing}, \bibinfo{address}{Sterling,
  VA, USA}.
\newblock
\showISBNx{978-1579224714}


\bibitem[\protect\citeauthoryear{Carver}{Carver}{2010}]%
        {Carver-RESER2010-ReportingGuidelines}
\bibfield{author}{\bibinfo{person}{Jeffrey~C Carver}.}
  \bibinfo{year}{2010}\natexlab{}.
\newblock \showarticletitle{Towards Reporting Guidelines for Experimental
  Replications: A Proposal}. In \bibinfo{booktitle}{\emph{1st international
  workshop on replication in empirical software engineering}},
  Vol.~\bibinfo{volume}{1}. Citeseer, \bibinfo{pages}{1--4}.
\newblock


\bibitem[\protect\citeauthoryear{Chattopadhyay and Chindaphone}{Chattopadhyay
  and Chindaphone}{2018}]%
        {chattopadhyay2018cocurricularpd}
\bibfield{author}{\bibinfo{person}{Ankur Chattopadhyay} {and}
  \bibinfo{person}{Bobby Chindaphone}.} \bibinfo{year}{2018}\natexlab{}.
\newblock \showarticletitle{A Nifty Inter-Class Peer Learning Model for
  Enhancing Student-Centered Computing Education, and for Generating Student
  Interests in Co-Curricular Professional Development}. In
  \bibinfo{booktitle}{\emph{2018 IEEE Frontiers in Education Conference
  (FIE)}}. \bibinfo{pages}{1--5}.
\newblock


\bibitem[\protect\citeauthoryear{Clear}{Clear}{2006}]%
        {Clear:2006:VCS:1315803.1315806}
\bibfield{author}{\bibinfo{person}{Tony Clear}.}
  \bibinfo{year}{2006}\natexlab{}.
\newblock \showarticletitle{Valuing Computer Science Education Research?}. In
  \bibinfo{booktitle}{\emph{Proceedings of the 6th Baltic Sea Conference on
  Computing Education Research: Koli Calling 2006}} (Uppsala, Sweden)
  \emph{(\bibinfo{series}{Baltic Sea '06})}. \bibinfo{publisher}{ACM},
  \bibinfo{address}{New York, NY, USA}, \bibinfo{pages}{8--18}.
\newblock
\urldef\tempurl%
\url{https://doi.org/10.1145/1315803.1315806}
\showDOI{\tempurl}


\bibitem[\protect\citeauthoryear{Clearinghouse}{Clearinghouse}{2020}]%
        {WWC2020}
\bibfield{author}{\bibinfo{person}{What~Works Clearinghouse}.}
  \bibinfo{year}{2020}\natexlab{}.
\newblock \bibinfo{title}{What Works Clearinghouse Standards Handbook, Version
  4.1}.
\newblock \bibinfo{howpublished}{\url{https://ies.ed.gov/ncee/wwc/Handbooks}}.
\newblock


\bibitem[\protect\citeauthoryear{Daniels and Pears}{Daniels and Pears}{2012}]%
        {daniels2012models}
\bibfield{author}{\bibinfo{person}{Mats Daniels} {and} \bibinfo{person}{Arnold
  Pears}.} \bibinfo{year}{2012}\natexlab{}.
\newblock \showarticletitle{Models and Methods for Computing Education
  Research}. In \bibinfo{booktitle}{\emph{Proceedings of the Fourteenth
  Australasian Computing Education Conference - Volume 123}} (Melbourne,
  Australia) \emph{(\bibinfo{series}{ACE '12})}. \bibinfo{publisher}{Australian
  Computer Society, Inc.}, \bibinfo{address}{AUS}, \bibinfo{pages}{95--102}.
\newblock
\showISBNx{9781921770043}


\bibitem[\protect\citeauthoryear{Edwards, Fulton, Holmes, Valentin, Beard, and
  Parker}{Edwards et~al\mbox{.}}{2018}]%
        {edwards2018syntax}
\bibfield{author}{\bibinfo{person}{John~M. Edwards}, \bibinfo{person}{Erika~K.
  Fulton}, \bibinfo{person}{Jonathan~D. Holmes}, \bibinfo{person}{Joseph~L.
  Valentin}, \bibinfo{person}{David~V. Beard}, {and} \bibinfo{person}{Kevin~R.
  Parker}.} \bibinfo{year}{2018}\natexlab{}.
\newblock \showarticletitle{Separation of syntax and problem solving in
  Introductory Computer Programming}. In \bibinfo{booktitle}{\emph{2018 IEEE
  Frontiers in Education Conference (FIE)}}. \bibinfo{pages}{1--5}.
\newblock


\bibitem[\protect\citeauthoryear{Fincher and Petre}{Fincher and Petre}{2004}]%
        {Fincher-Petre:04}
\bibfield{author}{\bibinfo{person}{Sally Fincher} {and} \bibinfo{person}{Marian
  Petre}.} \bibinfo{year}{2004}\natexlab{}.
\newblock \bibinfo{booktitle}{\emph{Computer Science Education Research}}.
\newblock \bibinfo{publisher}{Taylor and Francis}, \bibinfo{address}{Lisse, The
  Netherlands}.
\newblock
\showISBNx{978-9026519697}


\bibitem[\protect\citeauthoryear{Ham and Myers}{Ham and Myers}{2019}]%
        {ham2019guidedinquiry}
\bibfield{author}{\bibinfo{person}{Yeajin Ham} {and} \bibinfo{person}{Brandon
  Myers}.} \bibinfo{year}{2019}\natexlab{}.
\newblock \showarticletitle{Supporting Guided Inquiry with Cooperative Learning
  in Computer Organization}. In \bibinfo{booktitle}{\emph{Proceedings of the
  50th ACM Technical Symposium on Computer Science Education}} (Minneapolis,
  MN, USA) \emph{(\bibinfo{series}{SIGCSE '19})}.
  \bibinfo{publisher}{Association for Computing Machinery},
  \bibinfo{address}{New York, NY, USA}, \bibinfo{pages}{273–279}.
\newblock
\showISBNx{9781450358903}
\urldef\tempurl%
\url{https://doi-org.prox.lib.ncsu.edu/10.1145/3287324.3287355}
\showURL{%
\tempurl}


\bibitem[\protect\citeauthoryear{Hao, Smith~IV, Iriumi, Tsikerdekis, and
  Ko}{Hao et~al\mbox{.}}{2019}]%
        {Hao:2019TOCE:Replication}
\bibfield{author}{\bibinfo{person}{Qiang Hao}, \bibinfo{person}{David~H.
  Smith~IV}, \bibinfo{person}{Naitra Iriumi}, \bibinfo{person}{Michail
  Tsikerdekis}, {and} \bibinfo{person}{Amy~J. Ko}.}
  \bibinfo{year}{2019}\natexlab{}.
\newblock \showarticletitle{A Systematic Investigation of Replications in
  Computing Education Research}.
\newblock \bibinfo{journal}{\emph{ACM Trans. Comput. Educ.}}
  \bibinfo{volume}{19}, \bibinfo{number}{4}, Article \bibinfo{articleno}{42}
  (\bibinfo{date}{Aug.} \bibinfo{year}{2019}), \bibinfo{numpages}{18}~pages.
\newblock
\showISSN{1946-6226}
\urldef\tempurl%
\url{http://doi.acm.org/10.1145/3345328}
\showURL{%
\tempurl}


\bibitem[\protect\citeauthoryear{Heckman, Carver, Sherriff, and
  Al-Zubidy}{Heckman et~al\mbox{.}}{2021}]%
        {heckman2021systematic}
\bibfield{author}{\bibinfo{person}{Sarah Heckman}, \bibinfo{person}{Jeffrey~C.
  Carver}, \bibinfo{person}{Mark Sherriff}, {and} \bibinfo{person}{Ahmed
  Al-Zubidy}.} \bibinfo{year}{2021}\natexlab{}.
\newblock \bibinfo{title}{A Systematic Literature Review of Empiricism and
  Norms of Reporting in Computing Education Research Literature}.
\newblock
\newblock
\showeprint[arxiv]{2107.01984}~[cs.CY]


\bibitem[\protect\citeauthoryear{Heckman, Zhang, P\'{e}rez-Qui\~{n}ones, and
  Hawthorne}{Heckman et~al\mbox{.}}{2018}]%
        {heckman2018SIGCSEpaper}
\bibfield{author}{\bibinfo{person}{Sarah Heckman}, \bibinfo{person}{Jian
  Zhang}, \bibinfo{person}{Manuel~A. P\'{e}rez-Qui\~{n}ones}, {and}
  \bibinfo{person}{Elizabeth~K. Hawthorne}.} \bibinfo{year}{2018}\natexlab{}.
\newblock \showarticletitle{What is a SIGCSE Symposium Paper?}
\newblock \bibinfo{journal}{\emph{SIGCSE Bull.}} \bibinfo{volume}{50},
  \bibinfo{number}{3} (\bibinfo{date}{July} \bibinfo{year}{2018}),
  \bibinfo{pages}{3}.
\newblock
\showISSN{0097-8418}
\urldef\tempurl%
\url{https://doi.org/10.1145/3243071.3243073}
\showURL{%
\tempurl}


\bibitem[\protect\citeauthoryear{Ihantola, Vihavainen, Ahadi, Butler,
  B{\"o}rstler, Edwards, Isohanni, Korhonen, Petersen, Rivers,
  et~al\mbox{.}}{Ihantola et~al\mbox{.}}{2015}]%
        {ihantola2015educational}
\bibfield{author}{\bibinfo{person}{Petri Ihantola}, \bibinfo{person}{Arto
  Vihavainen}, \bibinfo{person}{Alireza Ahadi}, \bibinfo{person}{Matthew
  Butler}, \bibinfo{person}{J{\"u}rgen B{\"o}rstler},
  \bibinfo{person}{Stephen~H Edwards}, \bibinfo{person}{Essi Isohanni},
  \bibinfo{person}{Ari Korhonen}, \bibinfo{person}{Andrew Petersen},
  \bibinfo{person}{Kelly Rivers}, {et~al\mbox{.}}}
  \bibinfo{year}{2015}\natexlab{}.
\newblock \showarticletitle{Educational Data Mining and Learning Analytics in
  Programming: Literature Review and Case Studies}.
\newblock \bibinfo{journal}{\emph{Proceedings of the 2015 ITiCSE Working Group
  Reports}} (\bibinfo{year}{2015}), \bibinfo{pages}{41--63}.
\newblock


\bibitem[\protect\citeauthoryear{Jedlitschka and Pfahl}{Jedlitschka and
  Pfahl}{2005}]%
        {jedlitschka2005reporting}
\bibfield{author}{\bibinfo{person}{Andreas Jedlitschka} {and}
  \bibinfo{person}{Dietmar Pfahl}.} \bibinfo{year}{2005}\natexlab{}.
\newblock \showarticletitle{Reporting Guidelines for Controlled Experiments in
  Software Engineering}. In \bibinfo{booktitle}{\emph{2005 International
  Symposium on Empirical Software Engineering, 2005.}} IEEE,
  \bibinfo{pages}{10--pp}.
\newblock


\bibitem[\protect\citeauthoryear{Kitchenham, Al-Khilidar, Babar, Berry, Cox,
  Keung, Kurniawati, Staples, Zhang, and Zhu}{Kitchenham et~al\mbox{.}}{2008}]%
        {kitchenham2008evaluating}
\bibfield{author}{\bibinfo{person}{Barbara Kitchenham}, \bibinfo{person}{Hiyam
  Al-Khilidar}, \bibinfo{person}{Muhammed~Ali Babar}, \bibinfo{person}{Mike
  Berry}, \bibinfo{person}{Karl Cox}, \bibinfo{person}{Jacky Keung},
  \bibinfo{person}{Felicia Kurniawati}, \bibinfo{person}{Mark Staples},
  \bibinfo{person}{He Zhang}, {and} \bibinfo{person}{Liming Zhu}.}
  \bibinfo{year}{2008}\natexlab{}.
\newblock \showarticletitle{Evaluating Guidelines for Reporting Empirical
  Software Engineering Studies}.
\newblock \bibinfo{journal}{\emph{Empirical Software Engineering}}
  \bibinfo{volume}{13}, \bibinfo{number}{1} (\bibinfo{year}{2008}),
  \bibinfo{pages}{97--121}.
\newblock


\bibitem[\protect\citeauthoryear{Lishinski, Good, Sands, and Yadav}{Lishinski
  et~al\mbox{.}}{2016}]%
        {lishinski2016methodological}
\bibfield{author}{\bibinfo{person}{Alex Lishinski}, \bibinfo{person}{Jon Good},
  \bibinfo{person}{Phil Sands}, {and} \bibinfo{person}{Aman Yadav}.}
  \bibinfo{year}{2016}\natexlab{}.
\newblock \showarticletitle{Methodological Rigor and Theoretical Foundations of
  CS Education Research}. In \bibinfo{booktitle}{\emph{Proceedings of the 2016
  ACM Conference on International Computing Education Research}}.
  \bibinfo{pages}{161--169}.
\newblock


\bibitem[\protect\citeauthoryear{Luxton-Reilly, Albluwi, Becker, Giannakos,
  Kumar, Ott, Paterson, Scott, Sheard, and Szabo}{Luxton-Reilly
  et~al\mbox{.}}{2018}]%
        {luxton2018introductory}
\bibfield{author}{\bibinfo{person}{Andrew Luxton-Reilly},
  \bibinfo{person}{Ibrahim Albluwi}, \bibinfo{person}{Brett~A Becker},
  \bibinfo{person}{Michail Giannakos}, \bibinfo{person}{Amruth~N Kumar},
  \bibinfo{person}{Linda Ott}, \bibinfo{person}{James Paterson},
  \bibinfo{person}{Michael~James Scott}, \bibinfo{person}{Judy Sheard}, {and}
  \bibinfo{person}{Claudia Szabo}.} \bibinfo{year}{2018}\natexlab{}.
\newblock \showarticletitle{Introductory Programming: A Systematic Literature
  Review}. In \bibinfo{booktitle}{\emph{Proceedings Companion of the 23rd
  Annual ACM Conference on Innovation and Technology in Computer Science
  Education}}. \bibinfo{pages}{55--106}.
\newblock


\bibitem[\protect\citeauthoryear{Malmi, Simon, Sheard, Bednarik, Helminen,
  Korhonen, Myller, Sorva, Taherkhani, et~al\mbox{.}}{Malmi
  et~al\mbox{.}}{2010}]%
        {malmi2010characterizing}
\bibfield{author}{\bibinfo{person}{Lauri Malmi}, \bibinfo{person}{Simon},
  \bibinfo{person}{Judy Sheard}, \bibinfo{person}{Roman Bednarik},
  \bibinfo{person}{Juha Helminen}, \bibinfo{person}{Ari Korhonen},
  \bibinfo{person}{Niko Myller}, \bibinfo{person}{Juha Sorva},
  \bibinfo{person}{Ahmad Taherkhani}, {et~al\mbox{.}}}
  \bibinfo{year}{2010}\natexlab{}.
\newblock \showarticletitle{Characterizing Research in Computing Education: A
  Preliminary Analysis of the Literature}. In
  \bibinfo{booktitle}{\emph{Proceedings of the Sixth International Workshop on
  Computing Education Research}}. ACM, \bibinfo{pages}{3--12}.
\newblock


\bibitem[\protect\citeauthoryear{Margulieux, Ketenci, and Decker}{Margulieux
  et~al\mbox{.}}{2019}]%
        {margulieux2019measurements}
\bibfield{author}{\bibinfo{person}{Lauren Margulieux},
  \bibinfo{person}{Tuba~Ayer Ketenci}, {and} \bibinfo{person}{Adrienne
  Decker}.} \bibinfo{year}{2019}\natexlab{}.
\newblock \showarticletitle{Review of Measurements Used in Computing Education
  Research and Suggestions for Increasing Standardization}.
\newblock \bibinfo{journal}{\emph{Computer Science Education}}
  \bibinfo{volume}{29}, \bibinfo{number}{1} (\bibinfo{year}{2019}),
  \bibinfo{pages}{49--78}.
\newblock


\bibitem[\protect\citeauthoryear{McGill}{McGill}{2019}]%
        {mcgill-koli2019-replication}
\bibfield{author}{\bibinfo{person}{Monica~M. McGill}.}
  \bibinfo{year}{2019}\natexlab{}.
\newblock \showarticletitle{Discovering Empirically-Based Best Practices in
  Computing Education Through Replication, Reproducibility, and Meta-Analysis
  Studies}. In \bibinfo{booktitle}{\emph{Proceedings of the 19th Koli Calling
  International Conference on Computing Education Research}} (Koli, Finland)
  \emph{(\bibinfo{series}{Koli Calling '19})}. \bibinfo{publisher}{Association
  for Computing Machinery}, \bibinfo{address}{New York, NY, USA}, Article
  \bibinfo{articleno}{7}, \bibinfo{numpages}{5}~pages.
\newblock
\showISBNx{9781450377157}
\urldef\tempurl%
\url{https://dx.doi.org/10.1145/3364510.3364528}
\showURL{%
\tempurl}


\bibitem[\protect\citeauthoryear{{McGill} and {Decker}}{{McGill} and
  {Decker}}{2018}]%
        {mcgill-fie2018-cer-repository}
\bibfield{author}{\bibinfo{person}{Monica~M. {McGill}} {and}
  \bibinfo{person}{Adrienne {Decker}}.} \bibinfo{year}{2018}\natexlab{}.
\newblock \showarticletitle{Defining Requirements for a Repository to Meet the
  Needs of K-12 Computer Science Educators, Researchers, and Evaluators}. In
  \bibinfo{booktitle}{\emph{2018 IEEE Frontiers in Education Conference
  (FIE)}}. \bibinfo{pages}{1--9}.
\newblock


\bibitem[\protect\citeauthoryear{McGill, Decker, and Abbott}{McGill
  et~al\mbox{.}}{2018}]%
        {mcgill2018improving}
\bibfield{author}{\bibinfo{person}{Monica~M McGill}, \bibinfo{person}{Adrienne
  Decker}, {and} \bibinfo{person}{Zachary Abbott}.}
  \bibinfo{year}{2018}\natexlab{}.
\newblock \showarticletitle{Improving Research and Experience Reports of
  Pre-College Computing Activities: A Gap Analysis}. In
  \bibinfo{booktitle}{\emph{Proceedings of the 49th ACM Technical Symposium on
  Computer Science Education}}. ACM, \bibinfo{pages}{964--969}.
\newblock


\bibitem[\protect\citeauthoryear{Mirza, Conrad, Lloyd, Matni, and Gatin}{Mirza
  et~al\mbox{.}}{2019}]%
        {mirza2019talitreview}
\bibfield{author}{\bibinfo{person}{Diba Mirza}, \bibinfo{person}{Phillip~T.
  Conrad}, \bibinfo{person}{Christian Lloyd}, \bibinfo{person}{Ziad Matni},
  {and} \bibinfo{person}{Arthur Gatin}.} \bibinfo{year}{2019}\natexlab{}.
\newblock \showarticletitle{Undergraduate Teaching Assistants in Computer
  Science: A Systematic Literature Review}. In
  \bibinfo{booktitle}{\emph{Proceedings of the 2019 ACM Conference on
  International Computing Education Research}} (Toronto ON, Canada)
  \emph{(\bibinfo{series}{ICER '19})}. \bibinfo{publisher}{Association for
  Computing Machinery}, \bibinfo{address}{New York, NY, USA},
  \bibinfo{pages}{31–40}.
\newblock
\showISBNx{9781450361859}
\urldef\tempurl%
\url{https://doi-org.prox.lib.ncsu.edu/10.1145/3291279.3339422}
\showURL{%
\tempurl}


\bibitem[\protect\citeauthoryear{Neumann and Linzmayer}{Neumann and
  Linzmayer}{2021}]%
        {neumann2021emotions}
\bibfield{author}{\bibinfo{person}{Marion Neumann} {and} \bibinfo{person}{Robin
  Linzmayer}.} \bibinfo{year}{2021}\natexlab{}.
\newblock \showarticletitle{Capturing Student Feedback and Emotions in Large
  Computing Courses: A Sentiment Analysis Approach}. In
  \bibinfo{booktitle}{\emph{Proceedings of the 52nd ACM Technical Symposium on
  Computer Science Education}}. \bibinfo{publisher}{Association for Computing
  Machinery}, \bibinfo{address}{New York, NY, USA}, \bibinfo{pages}{541–547}.
\newblock
\showISBNx{9781450380621}
\urldef\tempurl%
\url{https://doi-org.prox.lib.ncsu.edu/10.1145/3408877.3432403}
\showURL{%
\tempurl}


\bibitem[\protect\citeauthoryear{Prather, Pettit, Becker, Denny, Loksa, Peters,
  Albrecht, and Masci}{Prather et~al\mbox{.}}{2019}]%
        {prather2019metacognition}
\bibfield{author}{\bibinfo{person}{James Prather}, \bibinfo{person}{Raymond
  Pettit}, \bibinfo{person}{Brett~A. Becker}, \bibinfo{person}{Paul Denny},
  \bibinfo{person}{Dastyni Loksa}, \bibinfo{person}{Alani Peters},
  \bibinfo{person}{Zachary Albrecht}, {and} \bibinfo{person}{Krista Masci}.}
  \bibinfo{year}{2019}\natexlab{}.
\newblock \showarticletitle{First Things First: Providing Metacognitive
  Scaffolding for Interpreting Problem Prompts}. In
  \bibinfo{booktitle}{\emph{Proceedings of the 50th ACM Technical Symposium on
  Computer Science Education}} (Minneapolis, MN, USA)
  \emph{(\bibinfo{series}{SIGCSE '19})}. \bibinfo{publisher}{Association for
  Computing Machinery}, \bibinfo{address}{New York, NY, USA},
  \bibinfo{pages}{531–537}.
\newblock
\showISBNx{9781450358903}
\urldef\tempurl%
\url{https://doi-org.prox.lib.ncsu.edu/10.1145/3287324.3287374}
\showURL{%
\tempurl}


\bibitem[\protect\citeauthoryear{Randolph, Julnes, Sutinen, and
  Lehman}{Randolph et~al\mbox{.}}{2008}]%
        {randolph2008methodological}
\bibfield{author}{\bibinfo{person}{Justus~J Randolph}, \bibinfo{person}{George
  Julnes}, \bibinfo{person}{Erkki Sutinen}, {and} \bibinfo{person}{Steve
  Lehman}.} \bibinfo{year}{2008}\natexlab{}.
\newblock \showarticletitle{A Methodological Review of Computer Science
  Education Research}.
\newblock \bibinfo{journal}{\emph{Journal of Information Technology Education:
  Research}}  \bibinfo{volume}{7} (\bibinfo{year}{2008}),
  \bibinfo{pages}{135--162}.
\newblock


\bibitem[\protect\citeauthoryear{Runeson and H{\"o}st}{Runeson and
  H{\"o}st}{2009}]%
        {runeson2009guidelines}
\bibfield{author}{\bibinfo{person}{Per Runeson} {and} \bibinfo{person}{Martin
  H{\"o}st}.} \bibinfo{year}{2009}\natexlab{}.
\newblock \showarticletitle{Guidelines for Conducting and Reporting Case Study
  Research in Software Engineering}.
\newblock \bibinfo{journal}{\emph{Empirical software engineering}}
  \bibinfo{volume}{14}, \bibinfo{number}{2} (\bibinfo{year}{2009}),
  \bibinfo{pages}{131}.
\newblock
\urldef\tempurl%
\url{https://doi.org/10.1007/s10664-008-9102-8}
\showDOI{\tempurl}


\bibitem[\protect\citeauthoryear{Schmidt}{Schmidt}{2009}]%
        {schmidt2009replication}
\bibfield{author}{\bibinfo{person}{Stefan Schmidt}.}
  \bibinfo{year}{2009}\natexlab{}.
\newblock \showarticletitle{Shall we Really do it Again? The Powerful Concept
  of Replication is Neglected in the Social Sciences}.
\newblock \bibinfo{journal}{\emph{Review of General Psychology}}
  \bibinfo{volume}{13}, \bibinfo{number}{2} (\bibinfo{year}{2009}),
  \bibinfo{pages}{90--100}.
\newblock
\urldef\tempurl%
\url{https://doi.org/10.1037/a0015108}
\showURL{%
\tempurl}


\bibitem[\protect\citeauthoryear{Schulz, Altman, and Moher}{Schulz
  et~al\mbox{.}}{2010}]%
        {Schulzc-BMJ2010-ConsortStatement}
\bibfield{author}{\bibinfo{person}{Kenneth~F Schulz},
  \bibinfo{person}{Douglas~G Altman}, {and} \bibinfo{person}{David Moher}.}
  \bibinfo{year}{2010}\natexlab{}.
\newblock \showarticletitle{CONSORT 2010 Statement: Updated Guidelines for
  Reporting Parallel Group Randomised Trials}.
\newblock \bibinfo{journal}{\emph{BMJ}}  \bibinfo{volume}{340}
  (\bibinfo{year}{2010}).
\newblock
\showISSN{0959-8138}
\urldef\tempurl%
\url{https://www.bmj.com/content/340/bmj.c332}
\showURL{%
\tempurl}


\bibitem[\protect\citeauthoryear{Shaw}{Shaw}{2003}]%
        {shaw2003writing}
\bibfield{author}{\bibinfo{person}{Mary Shaw}.}
  \bibinfo{year}{2003}\natexlab{}.
\newblock \showarticletitle{Writing good software engineering research papers}.
  In \bibinfo{booktitle}{\emph{Software Engineering, 2003. Proceedings. 25th
  International Conference on}}. IEEE, \bibinfo{pages}{726--736}.
\newblock


\bibitem[\protect\citeauthoryear{Sheard, Simon, Hamilton, and
  L{\"o}nnberg}{Sheard et~al\mbox{.}}{2009}]%
        {sheard2009analysis}
\bibfield{author}{\bibinfo{person}{Judy Sheard}, \bibinfo{person}{Simon},
  \bibinfo{person}{Margaret Hamilton}, {and} \bibinfo{person}{Jan
  L{\"o}nnberg}.} \bibinfo{year}{2009}\natexlab{}.
\newblock \showarticletitle{Analysis of Research into the Teaching and Learning
  of Programming}. In \bibinfo{booktitle}{\emph{Proceedings of the Fifth
  International Workshop on Computing Education Research}}. ACM,
  \bibinfo{pages}{93--104}.
\newblock


\bibitem[\protect\citeauthoryear{Stephens-Martinez}{Stephens-Martinez}{2021}]%
        {stephensmartinez2021studytactics}
\bibfield{author}{\bibinfo{person}{Kristin Stephens-Martinez}.}
  \bibinfo{year}{2021}\natexlab{}.
\newblock \showarticletitle{A Study of the Relationship Between a CS1 Student's
  Gender and Performance Versus Gauging Understanding and Study Tactics}. In
  \bibinfo{booktitle}{\emph{Proceedings of the 52nd ACM Technical Symposium on
  Computer Science Education}}. \bibinfo{publisher}{Association for Computing
  Machinery}, \bibinfo{address}{New York, NY, USA}, \bibinfo{pages}{679–685}.
\newblock
\showISBNx{9781450380621}
\urldef\tempurl%
\url{https://doi-org.prox.lib.ncsu.edu/10.1145/3408877.3432365}
\showURL{%
\tempurl}


\bibitem[\protect\citeauthoryear{Valentine}{Valentine}{2004}]%
        {valentine2004cs}
\bibfield{author}{\bibinfo{person}{David~W Valentine}.}
  \bibinfo{year}{2004}\natexlab{}.
\newblock \showarticletitle{CS Educational Research: A Meta-Analysis of SIGCSE
  Technical Symposium Proceedings}.
\newblock \bibinfo{journal}{\emph{ACM SIGCSE Bulletin}} \bibinfo{volume}{36},
  \bibinfo{number}{1} (\bibinfo{year}{2004}), \bibinfo{pages}{255--259}.
\newblock


\end{thebibliography}
